\documentclass[a4paper]{llncs}
\usepackage{listings}
\usepackage{hhline}
\usepackage{url} 
\usepackage{amsmath}
\usepackage{amssymb}
\usepackage[autolanguage]{numprint}
\usepackage{alltt}
\usepackage{graphicx}
\usepackage[ruled,linesnumbered]{algorithm2e}
\usepackage{epsfig}

\usepackage[normalem]{ulem}
\usepackage{etextools}
\usepackage{color}
\newcommand{\olivier}[2]{\ifempty{#1}{\textcolor{black}{#2}}{(\sout{#1}) \textcolor{red}{#2}}}

\newcommand{\eg}{e.g.}
\newcommand{\ie}{i.e.}
\newcommand{\interff}[2]{\ensuremath{[#1 , #2]}}
\newcommand{\interoo}[2]{\ensuremath{(#1 , #2)}}
\newcommand{\interfo}[2]{\ensuremath{[#1 , #2)}}
\newcommand{\interof}[2]{\ensuremath{(#1 , #2]}}

\newcommand{\CBMC}{\textsc{CBMC}}
\newcommand{\IEEESTD}{\textsc{IEEE}~754}
\newcommand{\FLUCTUAT}{\textsc{Fluctuat}}
\newcommand{\CDFL}{\textsc{CDFL}}

\newcommand{\FPCS}{\textsc{FPCS}}

\newcommand{\RAICP}{\textsc{rAiCp}}
\newcommand{\REALPAVER}{\textsc{RealPaver}}
\newcommand{\CPBPV}{\textsc{CPBPV}}

\newcommand{\OUR}{\textsc{CPBPV\_FP}}

\newcommand{\R}{\ensuremath{\mathbb{R}}}
\newcommand{\F}{\ensuremath{\mathbb{F}}}

\newcommand{\xrlb}{\underline{x}_{\scriptscriptstyle\R}}
\newcommand{\xrub}{\overline{x}_{\scriptscriptstyle\R}}

\newcommand{\xflb}{\underline{x}_{\scriptscriptstyle\F}}
\newcommand{\xfub}{\overline{x}_{\scriptscriptstyle\F}}

\urldef{\mails}\path|firstname.lastname@unice.fr|

\title{Searching critical values for floating-point programs\thanks{This work was partially supported by
      ANR COVERIF (ANR-15-CE25-0002)
}}

\author{H{\'e}l{\`e}ne Collavizza \and   Claude Michel \and Michel Rueher}
\institute{University of Nice--Sophia Antipolis, I3S/CNRS\\
BP 121, 06903 Sophia Antipolis Cedex, France\\
 \email{firstname.lastname@unice.fr}
}
\begin{document}

%
\lstset{language=C, frame=tb, commentstyle=\textrm, escapeinside={(*@}{@*)},%
        floatplacement=htb, basicstyle=\small\ttfamily, commentstyle=\rmfamily,%
        columns=flexible}
\npthousandsep{\,}
\npthousandthpartsep{}


\maketitle

\begin{abstract}

Programs with floating-point computations are often derived from
mathematical models or designed with the semantics of the real numbers
in mind. However, for a given input, the computed path with
floating-point numbers may significantly differ from the path
corresponding to the same computation with real numbers. As a
consequence, developers do not know whether the program can
actually produce very unexpected outputs. We introduce here a new constraint-based
approach that searches for test cases in the part of the
over-approximation where errors due to floating-point arithmetic could
lead to unexpected decisions.

\end{abstract}

\vspace*{-0.6cm}
\section{Introduction}\label{intro}

\vspace*{-0.2cm}

In numerous applications, programs with floating-point computations are 
derived from mathematical models over the real numbers. However,
computations on floating-point numbers are different from calculations in an
 idealised semantics\footnote{That's to say,  computations as close as possible to the mathematical semantics of the real numbers; for instance,  computations with arbitrary precision or computer algebra systems.} of real numbers~\cite{goldberg91}.
For some values of the input variables, the result of a sequence of
operations over the floating-point numbers can be significantly
different from the result of the corresponding
mathematical operations over the real numbers. 
As a consequence, the computed path with floating-point numbers may
differ from the path corresponding to the same computation with real
numbers. This  can entail wrong outputs and dangerous decisions of  critical systems.  That's why identifying these values is a crucial issue for programs controlling critical systems.

Abstract interpretation based error
analysis~\cite{Chen2008} of
finite precision implementations computes an over-approximation of the
errors due to floating-point operations. 
The point is that state-of-the-art
tools~\cite{delmasGoubaultPutotSouyrisTekkalVedrine09}
may generate numerous false alarms. In \cite{ponsiniMichelRueher14}, we
introduced a hybrid approach combining abstract interpretation
and constraint programming techniques that reduces the number of false
alarms. However, the remaining false alarms are very embarrassing since
we cannot know  whether the predicted unstable behaviors will occur with actual data.

More formally, consider a program $P$, a set of intervals $I$ defining
the expected input values of $P$, and an output variable $x$ of $P$
on which depend critical decisions, e.g., activating
an anti-lock braking system. Let $\interff{\xrlb}{\xrub}$ be   a
sharp approximation over the set of real numbers $\R$ of the domain of variable $x$ for any  input of
$P$. $\interff{\xflb}{\xfub}$  stands for the domain of variable $x$ in
the over-approximation computed over the set of floating-point $\F$
for input values of $I$. The range $\interff{\xrlb}{\xrub}$ can
be determined by calculation or from physical limits. It  includes a small tolerance to take
into account approximation errors, \eg{} measurement, statistical, or
even floating-point arithmetic errors. This tolerance -- specified by the user-- defines an acceptable
loss of accuracy between the value computed over the floating-point
numbers and the value calculated over the real numbers. Values outside the
interval $\interff{\xrlb}{\xrub}$ can lead a
program to misbehave, \eg{} take a wrong branch in the control flow.
%
  
The problem we address in this paper consists of verifying whether
there exist {\em critical values} in $I$ for which the
program can actually produce a result value of $x$  inside the suspicious intervals
$\interfo{\xflb}{\xrlb}$ and
$\interof{\xrub}{\xfub}$.  To handle this problem, we introduce a
new constraint-based approach that searches for test cases that hit the
suspicious intervals in programs with floating-point computations.
In other words, our framework reduces this test case generation
problem to a constraint-solving problem over the floating-point
numbers where the domain of a critical decision variable has been shrunk to
a suspicious interval.  
A constraint solver --based on filtering
techniques designed to handle constraints over floating-point
numbers-- is used to search values for the input data.
Preliminary results of experiments  on small programs with
classical floating-point errors are encouraging.

The  \OUR{}, the system we developed,  outperforms generate and test methods for programs with more than one input variable.
Moreover, these search strategies can prove in many cases that no critical value exists.

\section{Motivating example}

Before going into the details, we illustrate our approach on a small
example. Assume we want to compute the area of a triangle from the
lengths of its sides $a$, $b$, and $c$ with Heron's formula:
$$\sqrt{s*(s-a)*(s-b)*(s-c)}$$
where $s = (a+b+c)/2.$
The C program in Fig.~\ref{lst:heron} implements this formula, when
$a$ is the longest side of the triangle.
\begin{figure}[t] 
\begin{lstlisting}[numbers=left, numbersep=-1em,%
                   numberstyle=\tiny]
  /* Pre-condition :  a (*@$\ge$@*) b and a (*@$\ge$@*) c */
  float heron(float a, float b, float c) { 
    float s, squared_area;
    squared_area = 0.0f;
    if (a <= b + c) { (*@\label{line:heron:test}@*)
      s = (a + b + c) / 2.0f;
      squared_area = s*(s-a)*(s-b)*(s-c);
    }
    return sqrt(squared_area);
  }
\end{lstlisting}
\caption{Heron}
\label{lst:heron}
\end{figure}

The test of line~\ref{line:heron:test} ensures that the given lengths form
a valid triangle. 

Now, suppose that the input domains are \mbox{\lstinline|a| $\in
  \interff{5}{10}$} and \mbox{\lstinline|b|, \lstinline|c| $\in
  \interff{0}{5}$}. Over the real numbers, \lstinline|s| is greater
than any of the sides of the triangle and \lstinline|squared_area|
cannot be negative. Moreover, \lstinline|squared_area| cannot be
greater than 156.25 over the real numbers since the triangle area is
maximized for a right triangle with $b=c=5$ and $a=5\sqrt{2}$.
However, these properties may not hold over the floating-point numbers
because absorption and cancellation phenomena can occur\footnote{Let's remind that absorption in an addition occurs when  adding
two numbers of very different order of magnitude, and the result is the
value of the biggest number, \ie{}, when $x+y$ with $y\neq0$ yields $x$. 
Cancellation occurs in $s-a$ when $s$ is so
  close to $a$ that the subtraction cancels most of the significant
  digits of $s$ and $a$.}.

Tools performing value analysis over the floating-point numbers
~\cite{delmasGoubaultPutotSouyrisTekkalVedrine09,ponsiniMichelRueher12}
approximate the domain of \lstinline|squared_area| to the interval
$\interff{-1262.21}{979.01}$. Since this domain is an
over-approximation, we do not know whether input values leading to
\lstinline|squared_area| $<0$ or \lstinline|squared_area| $>156.25$
actually exist. Note that input domains --here a $\in$ [5,10] and b, c $\in$ [0,5]-- are usually provided by the user. 

Assume the value of the tolerance\footnote{Note that even this small
  tolerance may lead to an exception in statement 9.}  $\varepsilon$
is $ 10^{-5}$, the suspicious intervals for \lstinline|squared_area|
are $\interfo{-1262.21}{-10^{-5}}$ and
$\interof{156.25001}{979.01}$. \OUR{}, the system we developed,
generated test cases for both intervals:
\begin{itemize}
\vspace{-0.4cm}
\item \lstinline|a| $= 5.517474$, \lstinline|b| $= 4.7105823$,
  \lstinline|c| $= 0.8068917$, and \lstinline|squared_area|
  \olivier{}{equals} $-1.0000001 \cdot 10^{-5}$;

\item \mbox{\lstinline|a| $= 7.072597$}, \mbox{\lstinline|b| $=$
  \lstinline|c| $= 5$}, and  \lstinline|squared_area| equals $
  156.25003$.
\end{itemize}

 \OUR{} could also  prove the absence of test cases for a  tolerance $\varepsilon = 10^{-3}$ with 
\mbox{\lstinline|squared_area| $> 156.25 + \varepsilon$}.

In order to limit the loss of accuracy due to
cancellation~\cite{goldberg91}, line 7 of Heron's program can be rewritten as follows:
\vspace{-0.2cm}\begin{small}
\begin{verbatim}
     squared_area = ((a+(b+c))*(c-(a-b))*(c+(a-b))*(a+(b-c)))/16.0f;
\end{verbatim}
\end{small}
%

%
\vspace{-0.2cm} 
\noindent However, there are still some problems with this optimized program. Indeed,  \OUR{} found  the test case  \mbox{\lstinline|a| $= 7.0755463$},
  \mbox{\lstinline|b| $= 4.350216$}, \mbox{\lstinline|c| $= 2.72533$},
  and \lstinline|squared_area| equals $-1.0000001 \cdot
  10^{-5}$ for   interval  $
    \interfo{-1262.21}{-10^{-5}}$ of \lstinline|squared_area|. 
There are no more problems in the interval $
  \interof{156.25001}{979.01}$ and  \OUR{} did prove it.

\section{Framework for generating test cases}
\label{sec:approach}

This section details the framework we designed to generate test cases
reaching suspicious intervals for a variable $x$ in a program $P$ with
floating-point computations.  

The kernel of our framework is \FPCS{}~\cite{michelRueherLebbah01,michel02,botellaGotliebMichel06,marreMichel10}, a solver for constraints over the floating-point numbers; that's to say  a symbolic execution approach for floating-point problems which combines interval propagation with explicit search for satisfiable floating-point assignments. \FPCS{} is used inside the \CPBPV{}
bounded model checking framework~\cite{collavizzaRueherVanHentenryck10}. \OUR{} is the adaptation of \CPBPV{} for generating test cases that hit the
suspicious intervals in programs with floating-point computations.

The inputs of \OUR{} are: $P$, an annotated program; a critical test $ct$
for variable $x$;  $\interfo{\xflb}{\xrlb}$ or
$\interof{\xrub}{\xfub}$, a suspicious interval for $x$. 
Annotations of $P$ specify the range of the input variables of $P$ as well as the suspicious interval for $x$. The latter assertion is just posted before the critical test $ct$.

To compute the suspicious interval for $x$, we approximate the domain
of $x$ over the real numbers by $\interff{\xrlb}{\xrub}$, and over the
floating-point numbers by $\interff{\xflb}{\xfub}$. These
approximations are computed with
\RAICP{}~\cite{ponsiniMichelRueher14}, a hybrid system that combines
abstract interpretation and constraint programming techniques in a
single static and automatic analysis. The current implementation of
\RAICP{} is based upon the abstract interpreter
\FLUCTUAT{}~\cite{delmasGoubaultPutotSouyrisTekkalVedrine09}, the constraint solver over the reals \REALPAVER{}~\cite{granvilliersBenhamou06} and \FPCS{}.
The suspicious intervals for variable $x$ are denoted
$\interfo{\xflb}{\xrlb}$ and
$\interof{\xrub}{\xfub}$. 

\OUR{} performs first some pre-processing: $P$ is transformed into DSA-like
form\footnote{DSA stands for Dynamic Single Assignment. In DSA-like form, all variables are assigned exactly
  once in each execution path}. If the program contains loops, \OUR{} unfolds loops $k$ times where $k$ is a user specified constant.  Loops are handled in  \CPBPV{} and \RAICP{} with standard unfolding and abstraction techniques\footnote{In bounded model checking, $k$ is usually increased  until a counter-example is found or until the number of time units is large enough for the application.
}. So, there are no more loops in the program when we start the constraint generation process.  Standard slicing operations are also performed to reduce the size of the control flow graph.

In a second step,  \OUR{} searches for executable paths reaching $ct$. For each of these paths, the collected constraints are sent to \FPCS{}, which solves the corresponding constraint systems over the floating point numbers. \FPCS{} returns either a  satisfiable instantiation of the input variables of $P$, or  $\emptyset$.\\
 As said before, \FPCS{}~\cite{michelRueherLebbah01,michel02,botellaGotliebMichel06,marreMichel10}
  is a constraint solver designed to solve a set of constraints over
  floating-point numbers without losing any solution.  It uses 
  $2B$-consistency along with projection functions adapted to
  floating-point arithmetic ~\cite{michel02,botellaGotliebMichel06} 
  to  filter constraints over the floating-point numbers.  \FPCS{}
  also provides stronger consistencies like $kB$-consistencies, which
  allow better filtering results.  

The search of solutions in constraint systems over floating numbers is trickier than the standard bisection-based search in constraint systems over intervals of real numbers. Thus, we have also implemented different strategies combining selection of specific points and pruning. Details on theses strategies are given in the experiments section.

\OUR{} ends up with one of the following results: 
\begin{itemize}
\item a test case proving that $P$ can produce a suspicious value for
  $x$; 
\item a proof that no test case reaching the suspicious interval
  can be generated: this is the case if the loops in $P$ cannot be
  unfolded beyond the bound $k$ (See~\cite{collavizzaRueherVanHentenryck10} for
    details on bounded unfolding) ; 
\item an inconclusive answer: no test case could be generated but the
  loops in $P$ could be unfolded beyond the bound $k$. In other words,
  the process is incomplete and we cannot conclude whether $P$ may
  produce a suspicious value.
\end{itemize}

\section{Preliminary experiments}
\label{sec:experiments}

We experimented with \OUR{} on  six small programs with cancellation
and absorption phenomena, two very common pitfalls of floating-point
arithmetic. The benchmarks are listed in the first two columns of table \ref{times}. 

First two benchmarks concern the {\tt heron} program and the {\tt optimized\_heron} program with the suspicious intervals described in the section  \ref{intro}. 

Program {\tt slope} (see Fig.~\ref{lst:derivative}) approximates the
derivative of the square function $f(x) = x^2$ at a given point
$x_0$. More precisely, it computes the slope of a nearby
secant line with a finite difference quotient: $f'(x_0) \approx \frac{f(x_0 + h) - f(x_0 - h)}{2h}$.
Over the real numbers, the smaller $h$ is, the more accurate the
formula is. For this function, the derivative is given by $f'(x) = 2x$ which
yields exactly $26$ for $x = 13$.  Over the floats,
\FLUCTUAT{}~\cite{delmasGoubaultPutotSouyrisTekkalVedrine09}
approximates the return value of the slope program to the interval
$\interff{0}{25943}$ when $h \in \interff{10^{-6}}{10^{-3}}$ and
$x_0=13$.

\begin{figure}[t]
\begin{lstlisting}[%numbers=left, 
numbersep=-1em,%
                   numberstyle=\tiny]
  float slope(float x0, float h) { 
    float x1 = x0 + h;  float x2 = x0 - h;
    float fx1 = x1*x1;  float fx2 = x2*x2;
    float res = (fx1 - fx2) / (2.0*h);
    return res;
  }
\end{lstlisting}
\caption{Approximation of the derivative of $x^2$ by a slope}
\label{lst:derivative}

\begin{lstlisting}[numbersep=-1em,%
                   numberstyle=\tiny]
  float polynomial(float a, float b, float c) { 
    float poly = (a*a + b + 1e-5f) * c;
    return poly;
  }
\end{lstlisting}
\caption{Computation of polynomial $(a^2+b+10^{-5})*c$}
\label{lst:polynomial}
\end{figure}

Program {\tt polynomial} in Fig.~\ref{lst:polynomial} illustrates an
absorption phenomenon. It computes the
polynomial $(a^2+b+10^{-5})*c$. For input domains $a \in \interff{10^3}{10^4}$, $b \in \interff{0}{1}$
and $c \in \interff{10^3}{10^4}$, the minimum value of the polynomial
over the real numbers is \olivier{}{equal to} $1000000000.01$.

 {\tt simple\_interpolator} and {\tt simple square} are two benches extracted from \cite{GoubaultP13}.
The first bench  computes an interpolator, affine by sub-intervals while the second is a rewrite of a square root function
used in an industrial context.

All experiments were done on an Intel Core 2
  Duo at 2.8 GHz with 4 GB of memory running 64-bit Linux.
We assume C programs handling \IEEESTD{} compliant floating-point arithmetic,
intended to be compiled with GCC without any optimization option and
run on a x86\_64 architecture managed by a 64-bit Linux operating
system. Rounding mode was to the nearest, i.e.,  where ties round to the nearest even digit in the required position.

\subsection{Strategies and solvers}

We run \OUR{} with the following search strategies for the \FPCS{} solver:

\begin{itemize}
\item \lstinline|std|: standard prune \& bisection-based search used in constraint-systems over intervals : splits the selected variable domain in two  domains of equal size;
\item \lstinline|fpc|: splits the domain of the selected variable in five  intervals: 
\begin{itemize}
\item Three degenerated intervals containing only a single floating point number: the smallest float $l$, the largest float $r$, and the mid-point $m$;
\item Two open intervals $\interoo{l}{m}$ and $\interoo{m}{r}$;
\end{itemize}
\item \lstinline|fp3s|: selects 3 degenerated intervals containing only a single floating point number: the smallest float $l$, the largest float $r$, and the  mid-point $m$.
Hence, \lstinline|fp3s| is an incomplete method that might miss some solutions.
\end{itemize}

For all these strategies, we select first the variables with the
largest domain and we perform a 3B$-$consistency filtering step
before starting the splitting process.

We compared \OUR{} with  \CBMC{}~\cite{clarkeKroeningLerda04} and
\CDFL{}~\cite{DHKT12}, two state-of-the-art software bounded model checkers based on SAT solvers that are  able
to deal with floating-point computations. 
We also run a simple generate \& test strategy: the program is run
with randomly generated input values and we test whether the result is
inside the suspicious interval. The process is stopped as soon as a
test case hitting the suspicious interval is found.

\subsection{ Results } 

Table \ref{times} reports the results for the other strategies and
solvers. Since strategy \lstinline|fpc3s| is incomplete, we indicate whether
a test case was found or not. Column \lstinline|s?| specifies whether a test case actually exists. Note that the computation times of \CBMC{} and \CDFL{}
include the pre-processing time for generating the constraint systems;
the pre-processing time required by \CPBPV{} is around 0.6s
but \CPBPV{} is a non-optimised system written in \lstinline|java|.

\begin{table*}[t]
\begin{center}
\begin {tabular}{|l||l||r|r|r|r|r|r|}
\hline
Name & Condition & \CDFL{} & \CBMC{} & \lstinline|std| & \lstinline|fpc| & \lstinline|fpc3s| & \lstinline|s?| \\
\hline
\hline
{\lstinline|heron|} & $area < 10^{-5}$  & 3.874s & 0.280s &  $>$ 180 & 0.705 &  0.022 (n) & y \\
 & $area > 156.25 + 10^{-5}$  & $>$ 180s & 34.512s &  22.323 & 7.804 & 0.083 (n) & y \\ 
\hline
{\lstinline|optimized_heron|}  & $area < 10^{-5}$  & 7.618s & 0.932s &  $>$ 180 & 0.148 & 0.022 (n) & y \\
   & $area > 156.25 + 10^{-5}$  & $>$ 180s & $>$ 180s &  8.988 &
                                                                     30.477
                                                 & 0.101 (n) & n \\

\hline 
{\lstinline|slope|} with & $dh < 26.0 - 1.0$ &  2.014s & 1.548s & 0.021 & 0.012 & 0.012 (y) & y\\
$\,\,h \in [10^{-6}, 10^{-3}]$  & $dh > 26.0 + 1.0$ & 1.599s & 0.653s & 0.055 & 0.011 & 0.011 (y) & y \\
 & $dh < 26.0 - 10.0$  & 0.715s & 1.108s & 0.006 & 0.006 & 0.007 (n) & n \\
 & $dh > 26.0 + 10.0$  & 1.025s & 1.080s & 0.006 & 0.006 & 0.006 (n) & n \\
\hline

{\lstinline|polynomial|}   & $r < 10^9 + $ & 0.170s & 0.295s &  0.022 &  0.006 & 0.006 (y) & y \\
 & \hfill$0.0099999904 - 10^{-3}$ & & & & & & \\
\hline
{\lstinline|simple_interpolator|}   & $res < -10^{-5}$ & 0.296s & 0.264s &  0.018 & 0.012 & 0.012 (y) & y \\\hline
 {\lstinline|simple_square|}  & $S > 1.453125$ & $--$ & 1.079s &  0.012 & 0.012 & 0.012 (n) & n \\
\hline 
\end {tabular}

\caption{Results of the different solvers and strategies on the benchmarks}
\label{times}
\end{center}

\end{table*}

\section{Discussion}
\subsection{ Results analysis } 
\label{sec:discussion}
The generate \& test strategy behaves quite well on programs with only
one input variable when a test case exists but it is unable to find
any test case for programs with more than one input variable.  More
precisely, it found a test case in less than 0.008s for the 6 suspicious intervals of program \lstinline|slope| and for program \lstinline|simple_interpolator|.
The generate \& test strategy  failed to find a test within 180s in
all other cases.Of course, this strategy cannot show that there is no test case reaching the suspicious interval; so, it is of little interest here.

Strategy \lstinline|fpc| is definitely the
most efficient and most robust one on all these benchmarks. Note that
\CBMC{} and \CDFL{} could neither handle the initial,  nor the
optimized version of program \lstinline|heron| in  a timeout of 20
minutes whereas \FPCS{} found solutions in a reasonable time.  

These preliminary results  are very encouraging: they show that {\OUR} is effective for generating test cases for suspicious values outside the range of
acceptable values on small programs with
classical floating-point errors. More importantly,  a strong point of  {\OUR} is definitely its refutation capabilities.

 Of course, experiments on more significant benchmarks and on real applications are still necessary to evaluate the full capabilities and limits of \OUR{}.

\subsection{Related and further work}


The goals of software bounded model checkers based on SAT solvers are close to our
approach.  The point is that SAT
solvers tend to be inefficient on these problems due to the size of
the domains of floating-point variables and the cost of bit-vector
operations~\cite{DHKT12}. \CDFL{}~\cite{DHKT12} tries to
\olivier{}{address} this issue \olivier{}{by} embedding an abstract
domain in the conflict driven clause learning algorithm of a SAT
solver. SAT solvers often use bitwise representations of numerical operations, which may be very expensive (e.g., thousands of variables for one equation in CDFL).  Brain and al~\cite{HallerGBK12,BrainDGHK13} have recently introduced  a bit-precise decision procedure for the theory of floating-point arithmetic. The core of their approach is a generalization of the conflict-driven clause-learning algorithm used in modern SAT solvers. Their technique is significantly faster than a bit-vector encoding approach.
Note that the constraint programming techniques used in our approach
are better suited to generate several test cases than these SAT-based
approaches. The advantage of CP is that it provides a uniform framework for representing and handling integers, real numbers and floats. 
A new abstract-interpretation based robustness analysis of finite
precision implementations has recently been
proposed~\cite{GoubaultP13}  for sound rounding error propagation in a
given path in presence of unstable tests.


A close connection between our floating-point solvers and the  two
above-mentioned approaches is certainly worth exploring.

A second direction for further work concerns  the integration of our constraint-based
approach with  new abstract-interpretation based robustness analysis of finite
precision implementations for sound rounding error propagation in a
given path in presence of unstable tests.


\bibliographystyle{plain}
\small
\bibliography{ictss16_sus}

\end{document}